\documentclass[12pt,a4paper]{article}

\usepackage[a4paper,margin=2.5cm]{geometry}

\usepackage{newtxtext,newtxmath} 

\usepackage{microtype}
\usepackage{setspace}
\usepackage{titlesec}
\titleformat{\section}
  {\normalfont\bfseries\fontsize{12}{14}\selectfont}{\thesection}{1em}{}
\titlespacing*{\section}{0pt}{1.2ex plus .2ex}{0.8ex}
\titleformat{\subsection}
  {\normalfont\itshape\fontsize{12}{14}\selectfont}{\thesubsection}{1em}{}
\titlespacing*{\subsection}{0pt}{1.0ex plus .2ex}{0.6ex}

\usepackage[font=bf,size=small,justification=centering]{caption}
\usepackage{graphicx}
\usepackage{booktabs}
\usepackage{siunitx} 

\usepackage[style=apa,backend=biber,sortcites=true,doi=false,isbn=false,url=false,eprint=false]{biblatex}
\DeclareLanguageMapping{english}{english-apa}
\AtEveryBibitem{
  \clearfield{note}
  \clearlist{note}

    \ifentrytype{misc}{\toggletrue{bbx:url}}{}%
  
}

\usepackage{titling}
\pretitle{\begin{center}\vspace*{-1em}\LARGE\bfseries} 
\posttitle{\par\vspace{0.5em}\end{center}}
\preauthor{\begin{center}\normalsize}
\postauthor{\par\end{center}}
\predate{}\postdate{}

\usepackage{authblk}

\usepackage[hidelinks]{hyperref}

\title{Forecasting Conceptual Diffusion in Science: The Case of Quantum Computing}

\author[1,4]{Thomas Maillart}
\author[1,4]{Thibaut Chataing}
\author[2]{David Dosu}
\author[3]{Paul Bagourd}
\author[3]{Julian Jang-Jaccard}
\author[3]{Alain Mermoud}

\affil[1]{\texttt{thomas.maillart@unige.ch}, \texttt{thibaut.chataing@unige.ch}\\
Geneva School of Economics and Management, University of Geneva, Geneva, Switzerland}

\affil[2]{
Open Quantum Institute, CERN, Geneva, Switzerland}

\affil[3]{\texttt{julian.jang-jaccard@armasuisse.ch}\\
armasuisse Science + Technology, Switzerland}

\affil[4]{Faculty of Medicine, University of Geneva, Geneva, Switzerland}

\date{} 

\begin{document}
\maketitle

\renewcommand{\thefootnote}{}%
\footnotetext{An earlier version of this work was presented at Global Tech Mining Conference (GTM) 2026 (submission~\#117). This is a revised and extended preprint.}%
\renewcommand{\thefootnote}{\arabic{footnote}}%

\begin{abstract}
\noindent \textbf{Abstract.}
Understanding and anticipating scientific change requires models that distinguish between endogenous consolidation and exogenous diffusion of scientific concepts. Using the quantum computing subtree of concepts in OpenAlex, we construct a temporally resolved concept co-occurrence network and track each concept pair through its upstream citation lineage and downstream diffusion. We train LightGBM models on distributional and diversity-aware features to predict four outcomes: endogenous reinforcement, exogenous diffusion, their ratio, and diffusion entropy. After controlling for overall publication growth of the scientific body, endogenous reinforcement proves largely unpredictable in the primary quantum-computing benchmark. In contrast, exogenous diffusion and entropy are strongly predictable ($R^2$ up to $0.78$) and are driven by upstream heterogeneity, citation breadth, and distributional dispersion, as shown by SHAP analyses; replications on robotics, advanced materials, and neuro implants confirm that exogenous diffusion remains the top-ranked target across fields ($R^2_{\text{test}} \approx 0.60$--$0.87$), while endogenous predictability rises markedly in neuro implants ($R^2_{\text{test}} = 0.83$), indicating that the quantum-computing asymmetry does not generalise uniformly. Case studies reveal that sharp entropy increases coincide with the opening of new conceptual frontiers, while entropy collapses signal technological convergence or paradigm displacement. These results demonstrate that conceptual diffusion is governed by stable structural regularities embedded in semantic and citation environments. By identifying early diversity-based signals of cross-domain uptake, the approach provides a scalable foundation for anticipatory scientometrics, technology foresight, and innovation-oriented policy analysis in rapidly evolving research fields.
\end{abstract}

\vspace{0.5cm}
\section{Introduction}
\label{sec:introduction}

Scientific progress unfolds through the continual recombination and diffusion of ideas (\cite{fleming_recombinant_2001}). Advances in large-scale bibliometrics and open knowledge graphs now make these processes observable at scale, enabling empirical analyses of how conceptual relationships emerge, consolidate, and propagate (\cite{gu_forecasting_2025}). Research in the field of science of science (\cite{sinatra_quantifying_2016}) shows that innovation is shaped by structured patterns of conceptual integration, collaboration, and cumulative growth rather than isolated breakthroughs (\cite{uzzi_atypical_2013, fortunato_science_2018}). Concepts interact and co-occur within an evolving knowledge ecosystem, producing the complex dynamics that characterize scientific change. OpenAlex (\cite{priem_openalex_2022}) and related resources (e.g., Dimensions AI, Semantic Scholar) support concept-level representations of science, where concepts act as semantic units around which research communities organize. Concept co-occurrence networks have proven effective for mapping scientific structure, identifying emerging ideas, and forecasting conceptual combinations (\cite{salatino_how_2017, gu_forecasting_2025}). Complementary studies of citation dynamics show that scientific impact follows heavy-tailed distributions shaped by cumulative advantage (\cite{wang_measuring_2008}). Co-occurrences thus encode both intellectual inheritance and creative expansion. Two coupled processes govern conceptual evolution: upstream citations (i.e., what works {\it are} cited) capture the intellectual lineage on which a concept pair builds, while downstream citations (i.e., what works {\it have} cited) reflect how its ideas diffuse and are reinterpreted over time by other researchers of various fields. Together, they determine whether a combination remains endogenous or generalizes across domains. Yet these upstream–downstream mechanisms have rarely been examined jointly, and even less at the semantic concept-pair level. A central insight from prior work is the role of diversity in enabling conceptual innovation. Input diversity, i.e., combinations of distant or heterogeneous ideas, enhances novelty (\cite{shi_surprising_2023, wu_large_2019}), while adoption diversity across research semantic domains measures the breadth of downstream uptake. At the same time, preferential attachment and proportional growth generate self-reinforcing visibility (\cite{maillart_empirical_2008}). This study bridges these perspectives by modelling the joint upstream and downstream influence dynamics of concept pairs in quantum computing. Using a temporally evolving concept co-occurrence network enriched with citation-derived features, we analyse how knowledge propagates across multiple horizons and develop predictive tools for identifying which conceptual linkages are likely to diffuse.

The remainder of the paper proceeds as follows. Section~\ref{sec:background} reviews related literature. Section~\ref{sec:model} introduces our modelling framework and our research hypotheses. Section~\ref{sec:methods} describes the data and predictive setup. Section~\ref{sec:results} presents results and SHAP-based interpretation, and Section~\ref{sec:discussion} discusses implications for scientific conceptual evolution and forecasting. Section~\ref{sec:conclusion} concludes. 

\vspace{0.5cm}
\section{Background}
\label{sec:background}

Scientific progress is widely recognised as a cumulative, networked process in which ideas evolve through structured interactions across conceptual, collaborative, and institutional systems (\cite{fleming_recombinant_2001, arthur_nature_2009, strumsky_complexity_2010}). Rather than emerging from isolated contributions, innovation arises from the organisation of scientific knowledge into interdependent communities and evolving knowledge structures (\cite{fortunato_science_2018, uzzi_atypical_2013, wagner_international_2019}). These structures exhibit characteristic statistical regularities, such as heavy-tailed activity, cumulative advantage, and proportional growth, which constrain both the emergence and the predictability of scientific impact (\cite{wang_measuring_2008, newman_structure_2001}).

Within these structures, diversity has emerged as a central driver of creativity and long-term influence. Studies of combinatorial innovation show that novel or atypical recombinations of distant ideas disproportionately generate influential advances (\cite{uzzi_atypical_2013, shi_surprising_2023, youn_invention_2015}). Input diversity (e.g., heterogeneity in referenced concepts, disciplines, or methods) supports recombinant search (\cite{arthur_nature_2009}) and predicts disruptive potential (\cite{enduri_does_2015, wu_large_2019}). Likewise, adoption diversity, i.e., the breadth of downstream uptake, signals whether ideas diffuse beyond their local communities and reshape multiple research areas (\cite{veugelers_scientific_2019, shi_surprising_2023}). Together, these insights position diversity as a key mechanism shaping how scientific ideas build from and propagate.

To study these processes, researchers increasingly model science at the semantic level, using concept co-occurrence networks in which nodes represent scientific concepts and edges mark their joint appearance in publications. These networks capture the combinatorial substrate of scientific discovery and reveal temporal patterns of emergence, consolidation, and obsolescence (\cite{kuhn_structure_1997,salatino_how_2017, chen_science_2017}). Upstream citations expose the intellectual lineage feeding conceptual combinations, while downstream citations trace how ideas diffuse and are reinterpreted across domains. Despite their relevance, upstream–downstream dynamics remain under-explored at the level of concept pairs, which is precisely where conceptual recombination is most explicit (\cite{pan_memory_2018}).

Recent advances in large-scale open bibliographic infrastructures have transformed the ability to analyse these semantic and citation structures at scale. OpenAlex, in particular, provides a comprehensive graph of works, citations, authors, institutions, and hierarchical concept ontologies (\cite{priem_openalex_2022}). The availability of these datasets has enabled data-driven forecasting approaches: machine-learning models trained on evolving knowledge graphs can identify emerging topics (\cite{percia_david_measuring_2023,dolamic_automated_2024}), anticipate conceptual linkages, and detect early signals of technological transitions (\cite{krenn_forecasting_2023, gu_forecasting_2025}). Deep-learning architectures further support citation prediction (\cite{mistele_predicting_2019, zhang_predicting_2024}), conceptual-evolution modelling (\cite{krenn_predicting_2020}), and domain-specific technology forecasting (\cite{gui_technology_2021, hu_technology_2022, li_technology_2022}). These developments collectively demonstrate that scientific diffusion is not random but governed by measurable regularities embedded in coupled semantic–citation environments.

Against this backdrop, quantum computing represents an interesting setting to study conceptual recombination: it is a frontier field marked by high interdisciplinarity, rapid knowledge turnover, and strong theoretical–technological co-evolution (\cite{preskill_quantum_2018}). By unifying upstream lineage and downstream diffusion in a temporally evolving concept co-occurrence network, our study aims to reveal structural mechanisms governing conceptual influence and to advance forecasting methods that support anticipatory analysis in fast-moving scientific domains.

\vspace{0.5cm}
\section{Model \& Hypotheses}
\label{sec:model}

Scientific knowledge can be represented as a temporally evolving semantic network in which concepts gain meaning through repeated co-appearance in scholarly work (\cite{fortunato_science_2018}). Co-occurrence frequencies trace shifts in attention and make concept pairs a tractable unit for analysing novelty, integration, and diffusion (\cite{uzzi_atypical_2013,salatino_how_2017}). Each pair evolves within a dual environment: an upstream structure capturing the intellectual inputs on which it draws, and a downstream structure reflecting how its ideas propagate. This view aligns with evolutionary perspectives emphasizing recombination across heterogeneous knowledge bases (\cite{rzhetsky_choosing_2015}). Formally, for each year $t$ we construct a weighted, undirected graph,
\begin{equation}
G_t = (V_t, E_t, w_t),
\end{equation}
where $V_t$ is the active concept set, $E_t$ the concept pairs, and $w_t$ their co-occurrence weights in number of publications per year carrying each pair. To ensure substantive pair weights, only concept pairs supported by at least five publications in one year are retained. For a focal pair, the upstream environment consists of all references cited in papers containing the pair at year $t$. From these cited works, we compute a feature vector summarizing the distribution of upstream concept-pair weights in year $t-1$, yielding a Markovian approximation of the conceptual neighbourhood from which the pair draws momentum. The downstream environment is defined by all works citing the focal pair publications over the subsequent five years. Concept pairs extracted from this citing corpus quantify the breadth and direction of diffusion, i.e., how widely and in what contexts the focal pair reappears with other pairs as research fields evolve. Together, the dynamic graph and upstream–downstream representations allow us to distinguish mechanisms of stability from mechanisms of dissemination and to evaluate whether conceptual momentum arises from internal reinforcement or from the diversity of ideas surrounding a pair. We advance two hypotheses reflecting these distinct mechanisms:

\paragraph{Hypothesis 1: Endogenous self-reinforcement reflects proportional growth.}  
Endogenous reinforcement -- citations originating from within the same concept pair -- tends to follow “rich-get-richer” proportional-growth processes (\cite{maillart_empirical_2008}). Indeed, as scientific output expands, established concept pairs attract increasingly more citations, exhibiting exponential growth dynamics that eventually bend toward a saturation regime or effective carrying capacity (\cite{percia_david_measuring_2023}). In quantum science, for instance, long-standing pairings such as \emph{superconducting qubits} and \emph{Josephson junctions} have generated dense self-referential citation loops over decades, driven largely by cumulative experimentation and incremental optimization rather than by cross-domain recombination. We therefore expect endogenous dynamics to predominantly track corpus-level growth through self-momentum, reflecting structural reinforcement rather than novel diffusion mechanisms.

\paragraph{Hypothesis 2: Diffusion is driven by diversity and heterogeneity.}  
Concept pairs nourished by diverse, information-rich upstream foundations diffuse more broadly across conceptual boundaries. High upstream entropy, heterogeneity, and dispersion are therefore expected to correlate with stronger exogenous uptake and greater downstream diversity, consistent with evidence that heterogeneous inputs facilitate the recombination of distant knowledge components (\cite{enduri_does_2015}), promote novelty and influence (\cite{wu_large_2019}), and enable broader reinterpretation across domains (\cite{shi_surprising_2023}). In quantum science, for example, early linkages between \emph{quantum error correction} and \emph{topological phases of matter} drew on a heterogeneous upstream literature spanning information theory, condensed-matter physics, and topology, and subsequently diffused into areas such as topological quantum computing, fault-tolerant architectures, and quantum materials; by contrast, concept pairs grounded in narrower upstream traditions tend to propagate within more specialized communities. Diffusion is thus expected to follow identifiable structural regularities rooted in upstream diversity rather than random expansion.\\

These two hypotheses link micro-level citation structures to macro-level trajectories of scientific change: proportional growth underpins self-momentum of concept pairs, whereas structured recombination drives conceptual reach and cross-domain influence.

\vspace{0.5cm}
\section{Methods \& Predictive Models}
\label{sec:methods}

\paragraph{Dataset construction.}
We constructed our dataset using OpenAlex, a large-scale bibliographic knowledge graph indexing scholarly works together with their conceptual annotations (\cite{priem_openalex_2022}). We extracted all publications associated with the hierarchical subtree rooted in the Level-3 concept \emph{Quantum Computer} (C58053490), retaining Level-4 and Level-5 concepts with a concept-score above $0.32$ (the OpenAlex-recommended threshold for reliable annotations). These include, for example, Level-4 concepts such as quantum error correction, quantum algorithms, and quantum hardware, and more granular Level-5 concepts such as topological qubits, superconducting qubits, surface codes, and variational quantum algorithms. Dates were normalised to calendar years because over $90\%$ of publications indexed in OpenAlex have only yearly temporal resolution. The resulting corpus (1990--2023) provides a consistent basis for examining how quantum-computing concepts combine with each other and with concepts outside the Quantum Computer subtree. To ensure substantive pair weights, only concept pairs supported by at least five publications in a given year are retained, and pairs are further stratified by annual paper counts to balance the representation of dominant and niche combinations.

\paragraph{Focal-year cohort and right-censoring.}
Predictive models in Table~\ref{tab:metrics_summary} use only \emph{censoring-free} focal years $t \in \{1996,\ldots,2018\}$, for which the full five-year downstream window $(t{+}1,\ldots,t{+}5)$ lies within the 2023 snapshot ($t{+}5 \le 2023$). Focal years $2019$--$2023$ remain in the bibliographic corpus for descriptive statistics but are excluded from the primary fit because downstream citation counts and entropies are right-censored (incomplete tails). Upstream features at $t{-}1$ are fully observed for all retained years. Appendix~\ref{app:censoring} documents cohort sizes and contrasts this protocol with the comparative validation holdout (focal years 2022--2023, Section~\ref{sec:cross_domain}).

\paragraph{Upstream and downstream citation environments.}
For each year $t$, we constructed weighted concept co-occurrence pairs from all publications containing each pair (\emph{focal papers}). We then assembled two complementary citation environments: (i) an \emph{upstream} environment consisting of all references cited by the focal papers at $t{-}1$, and (ii) a \emph{downstream} environment comprising all papers citing them over the next five years ($t{+}1, \ldots, t{+}5$). The five-year downstream horizon is motivated by three considerations. First, citation accrual in scholarly publications is heavy-tailed but reaches a substantial share of its long-run total within roughly five years of publication; empirical work on bibliometric forecasting uses windows ranging from 2--3 years \cite{min_predicting_2021} to 6 years, with five years a common compromise that captures both early uptake and the onset of cross-domain diffusion. Second, a five-year window balances coverage against right-censoring induced by the OpenAlex snapshot date: shorter windows (e.g., 3 years) discard informative signal for older cohorts, whereas longer windows (e.g., 7 years) further restrict the set of focal years for which the full window is observed (right-censoring of focal years from 2019 onward is discussed in Section~\ref{sec:discussion}). Third, the choice is consistent with the time horizons used in adjacent breakthrough-prediction work on OpenAlex-style data. Sensitivity to alternative downstream windows (3 and 7 years) is a useful robustness check; we do not expect the qualitative conclusions to depend on the exact choice.

\paragraph{Upstream feature construction.}
Upstream construction involved extracting all concept pairs (above the $0.32$ threshold) weighted by their occurrences in the cited works at $t{-}1$. This yields a Markovian approximation of the conceptual lineage supporting the focal pair, through its own focal pair weight relative to the weight of other pairs found in the cited papers. From the resulting distribution of normalised concept-pair weights (counts divided by annual totals), we computed $28$ statistical features grouped into seven families summarised in Table~\ref{tab:features}. The choice of these families follows established practice in the analysis of heavy-tailed bibliometric distributions: quantile- and moment-based descriptors are standard summaries of skewed empirical distributions; \emph{Shannon entropy} (\cite{jost_entropy_2006}) quantifies the diversity of upstream conceptual support; geometric and harmonic central-tendency measures are robust under multiplicative and heavy-tailed regimes typical of citation data (\cite{maillart_empirical_2008,saichev_theory_2009}); and the explicit endogenous/exogenous decomposition mirrors the persistence-versus-recombination distinction documented in science-of-science work (\cite{uzzi_atypical_2013,wu_large_2019,shi_surprising_2023}). Together, these features capture low-order and higher-order distributional signals -- asymmetry, heavy-tail behaviour, and the balance between internal reinforcement and external uptake -- without committing to a parametric distributional form.

\begin{table}[htbp]
\caption{Upstream features ($28$ indicators) grouped by family, with definitions and grounding references.}
\small
\begin{tabular}{p{3.0cm}p{2.8cm}p{6.5cm}p{2.6cm}}
\toprule
Family & Indicators & Definition / role & References \\
\midrule
Quantiles & $q_{05}, q_{25}, q_{50}, q_{75}, q_{95}$ & Empirical quantiles of normalised pair-weight distribution; non-parametric tail descriptors. & standard \\
Central tendency & arithmetic, geometric, harmonic mean; median; root mean square (RMS) & Location summaries; geometric/harmonic means are robust under multiplicative, heavy-tailed regimes. & \cite{maillart_empirical_2008,saichev_theory_2009} \\
Dispersion & standard deviation (SD), geometric standard deviation (geometric SD), median absolute deviation (MAD), min, max & Spread of upstream support; geometric SD is the natural multiplicative analogue of SD for log-normal-like data. & \cite{saichev_theory_2009} \\
Shape & skewness, kurtosis & Asymmetry and tail-heaviness of the upstream distribution. & standard \\
Diversity & Shannon entropy & Diversity of upstream pair support (in nats); core diffusion-breadth indicator. & \cite{jost_entropy_2006} \\
Volume & total normalised weight, \# unique cited papers & Aggregate intellectual ``mass'' supporting the focal pair. & --- \\
Endo / exo split & endogenous count, exogenous count & Decomposition into self-reinforcement vs.\ cross-pair recombination, mirroring the persistence-vs-diffusion distinction. & \cite{uzzi_atypical_2013,wu_large_2019,shi_surprising_2023} \\
\bottomrule
\end{tabular}
\label{tab:features}
\end{table}

Downstream construction involved extracting concept pairs present in the citing corpus, capturing how the focal pair is reinterpreted and integrated into new contexts (together with other concept pairs) over the period ($t{+}1$, \ldots, $t{+}5$). Together, these upstream and downstream flows formalise how concept pairs accumulate intellectual inputs and propagate influence across conceptual space without re-describing the co-occurrence process already defined above. We defined four complementary target variables capturing key dimensions of future citation behaviour: (i) {\it endogenous} weight of focal concepts across citing papers in the next five years; (ii) {\it exogenous} weight of other concepts across citing papers; the (iii) {\it endogenous citation ratio} (with a small $\epsilon$ to avoid division by zero), and (iv) the {\it Shannon entropy} of concept pair weight distribution. These outcomes correspond to internal reinforcement, cross-domain diffusion, relative balance between the two, and the breadth of interpretive contexts. They also map into the above formulated hypotheses (Section \ref{sec:model}). 

Considering data splitting, concept pair weight distributions (downstream) typically exhibit heavy-tailed, highly skewed characteristics, with many concept pairs receiving few citations and a small fraction receiving disproportionately high attention. To ensure representative train-validation splits that preserve this distributional structure, we implemented a stratified sampling strategy based on logarithmic binning: Zero-valued observations were assigned to dedicated strata (predicted weight is $0$); positive target values were binned logarithmically, with bin counts adjusted to ensure at least two points per stratum. We then performed an 80–20 split using \textit{train-test-split} with stratification to preserve empirical distributional structure. Regression models were fitted using \textit{LightGBM} (\cite{ke_lightgbm_2017}), chosen for its efficiency and ability to capture nonlinear feature interactions. Hyperparameters were tuned via Bayesian optimisation using Optuna (\cite{akiba_optuna_2019}) with the Tree-Structured Parzen Estimator sampler, running 20 trials per target variable to maximise validation $R^2$. The search space spanned tree depth, number of leaves, learning rate, number of estimators, sampling ratios, and L1/L2 regularisation terms. Performance was evaluated using $R^2$, mean absolute error (MAE), and root mean squared logarithmic error (RMSLE) to capture variance explained, average magnitude of errors, and proportional accuracy across skewed citation regimes. Model interpretability was achieved via Shapley additive explanations (SHAP) (\cite{lundberg_unified_2017}), using TreeSHAP (\cite{lundberg_local_2020}) to compute exact feature attributions for gradient-boosted trees. SHAP decomposes each prediction into additive feature contributions, enabling direct interpretation of which distributional characteristics, such as entropy, dispersion, or tail weight, drive model behaviour. All analyses were implemented in Python using \texttt{pandas}, \texttt{numpy}, \texttt{scipy.stats}, \texttt{scikit-learn}, \texttt{lightgbm}, \texttt{optuna}, and \texttt{shap}, with fixed random seeds to ensure reproducibility.

\paragraph{Comparative validation protocol.}
To assess whether the diffusion asymmetries reported for quantum computing generalise beyond a single testbed, we replicated the same forecasting protocol on three additional OpenAlex subtrees: robotics, advanced materials, and neuro implants (Appendix~\ref{app:domains}). All comparative runs used the same OpenAlex validation subsample ($\approx 40\%$ of snapshot works by partition volume, structurally representative across 1990--2024). Focal years 2022--2023 were held out for testing; earlier years were used for training with fixed LightGBM hyperparameters (no per-domain Optuna). Because the 2023 snapshot does not yet contain the full five-year citation tail for these recent focal years, comparative test metrics reflect \emph{partial} downstream windows (Appendix~\ref{app:censoring_compare}), unlike Table~\ref{tab:metrics_summary}, which restricts to censoring-free focal years 1996--2018 with complete horizons. Metrics are \emph{within-domain} replications, not train-on-one-field / test-on-another experiments.

\vspace{0.5cm}
\section{Results}
\label{sec:results}

In this section, we examine how well the predictive model captures key dimensions of conceptual propagation in science, ranging from endogenous reinforcement to exogenous diffusion. By comparing performance across multiple tasks and interpreting feature contributions through SHAP analysis, we assess not only the quantitative accuracy of the model but also the underlying structural mechanics that govern how scientific ideas consolidate and spread over time.

\subsection{Analysis and Interpretation of Citation Variance Prediction Results}
\label{sec:results_interpretation}

Table~\ref{tab:metrics_summary} reports predictive performance on the censoring-free cohort (Methods, Appendix~\ref{app:censoring}) across four related tasks that capture complementary dimensions of conceptual propagation: endogenous count (self-reinforcement within the focal pair's downstream lineage), exogenous count (citations by papers with other concept pairs), the endo/exo ratio, and entropy (diffusion diversity). The reported metrics -- $R^2$, mean absolute error (MAE), and root mean squared logarithmic error (RMSLE) -- are evaluated on both training and held-out test sets. Overall, the model shows consistent generalisation, with only a modest drop from training to test performance, indicating appropriate regularisation and limited overfitting. Yet, predictive strength varies markedly across tasks, revealing asymmetries in the underlying conceptual dynamics. The model exhibits almost no explanatory power for endogenous citation activity ($R^2_{\text{test}} = 0.0179$), suggesting that self-referential reinforcement within a focal concept pair own lineage is largely unpredictable from available features. Because the model operates on normalized data, this does not imply no endogenous growth {\it per se}: It means that endogenous growth goes overall along with the global exponential growth trend of scientific output (\cite{percia_david_measuring_2023}) with any remaining signal reflecting local idiosyncracy rather than systematic self-reinforcing processes. In contrast, the exogenous count task achieves strong and consistent predictive power ($R^2_{\text{test}} = 0.78$). This indicates the exogenous diffusion of concepts -- propagating into papers that blend other concept pairs -- is governed by stable and measurable relationships between upstream knowledge structures and current co-occurrence patterns. Even after adjusting for overall growth, these patterns persist, showing that conceptual diffusion (i.e., a focal concept pair being cited by papers without this focal concept pair) is an inherently structured and content-driven phenomenon rather than a simple by-product of publication volume.

\begin{table}[htbp]
\caption{Citation variance prediction performance (primary benchmark). Focal years 1996--2018 only ($n = 6\,978$ pair--years): each row has a complete five-year downstream window in the 2023 OpenAlex snapshot. Growth-normalised targets; stratified 80/20 train--test split; Optuna-tuned LightGBM.}
\small
\begin{tabular}{lrrrrrr}
\toprule
Task & Train R² & Test R² & Train MAE & Test MAE & Train RMSLE & Test RMSLE \\
\midrule
Endo. count & 0.0218 & 0.0179 & 0.0000 & 0.0000 & 0.0000 & 0.0000 \\
Exo. count & 0.8364 & 0.7804 & 0.0012 & 0.0014 & 0.0026 & 0.0028 \\
Ratio  Endo/Exo & 0.6457 & 0.4715 & 0.0005 & 0.0006 & 0.0007 & 0.0008 \\
Entropy & 0.7650 & 0.6866 & 0.3120 & 0.3602 & 0.0467 & 0.0530 \\
\bottomrule
\end{tabular}
\label{tab:metrics_summary}
\end{table}

The ratio of endogenous to exogenous activity shows intermediate predictability ($R^2_{\text{test}} = 0.47$). This suggests that the model captures regularities in how concepts balance internal consolidation and outward influence. With normalisation, this ratio becomes a scale-independent measure of a concept pair's functional role in the evolving scientific landscape, highlighting transitions between specialised, inwardly focused dynamics and broader, integrative ones. Entropy reinforces this picture with a comparably high predictive accuracy ($R^2_{\text{test}} = 0.69$). The model effectively anticipates how broadly a concept pair's influence radiates across other research areas. Because entropy is intrinsically scale-free, its predictability confirms that the diversity of conceptual diffusion depends mainly on semantic proximity, upstream connectivity, and structural embedding rather than on absolute citation magnitude.

These findings show that once the evolution of scientific production is normalised out, only exogenous diffusion and diversity-related processes retain strong predictive signals. The endogenous, self-referential dynamics of concepts appear largely proportional to the size evolution of the research corpus, whereas cross-domain propagation reflects deeper, reproducible mechanisms of knowledge evolution.

\begin{figure}[h!]
    \centering
    \includegraphics[width=1\linewidth]{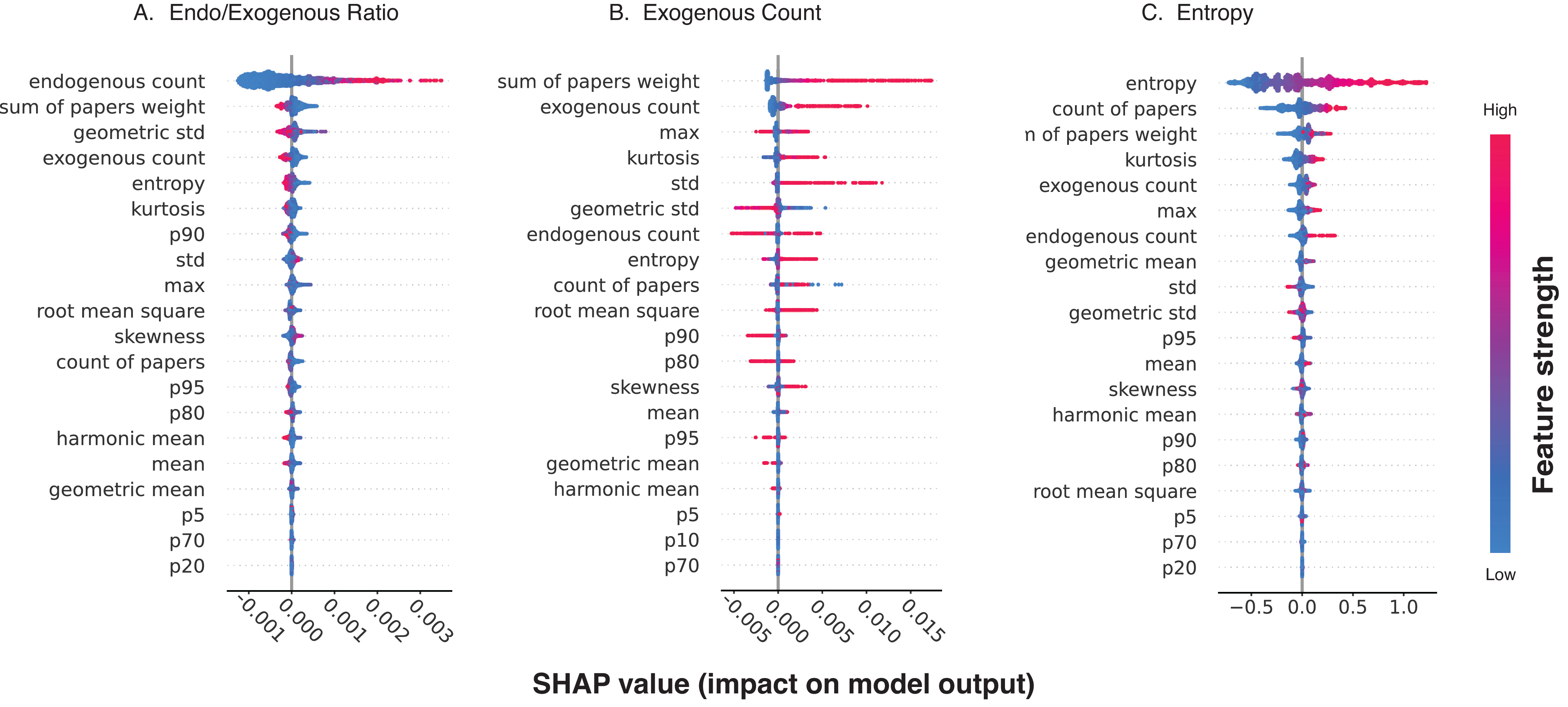}
    \caption{SHAP summary plots for {\bf A.} Endo/Exogenous Ratio, {\bf B.} Exogenous Count, and {\bf C.} Entropy. Features are ranked by mean absolute SHAP value, with colour indicating feature magnitude. High citation volume, diversity, and dispersion consistently drive diffusion-related outcomes, revealing a dominant regime of outward conceptual radiation fuelled by heterogeneity.}
    \label{fig:shap_value}
\end{figure}

\subsection{SHAP Results Interpretation}
Figure~\ref{fig:shap_value} presents SHAP analyses for the three informative prediction tasks: (i) endo/exogenous ratio, (ii) exogenous count, and (iii) entropy. It ranks features by their mean absolute contributions. The results highlight which statistical properties of citation distributions most strongly shape diffusion outcomes. For the {\it Endo/Exogenous Ratio}, the leading features are endogenous count, total cited-paper weight, and geometric standard deviation. Higher endogenous count increases the ratio, signalling greater internal consolidation, although overall effects remain modest. In contrast, larger cited-paper weights and greater geometric variability decrease the ratio, indicating that concept pairs grounded in heterogeneous and well-cited foundations tend to diffuse outward rather than reinforce internally. Secondary features similarly suggest that endogenous focused concept pairs exhibit narrower and less diverse citation structures. For the {\it Exogenous Count}, the dominant predictors are the total normalized citation count (\textit{sum of papers weight}) and exogenous count. High values of both substantially increase predicted diffusion, confirming that pairs supported by diverse and widely cited intellectual lineages propagate more effectively across research domains. Dispersion-related measures (standard deviation, kurtosis) also contribute positively, showing that moderately diverse upstream citation environments facilitate cross-domain uptake. For {\it Entropy}, the prediction is driven primarily by the entropy of concept-pairs in upstream citations, the number of cited papers, and total cited-paper weight, with kurtosis again contributing positively. High entropy and broad citing bases increase predicted diffusion breadth, indicating that concept pairs embedded in rich and heterogeneous knowledge environments generate wider and more balanced downstream influence.

Across all tasks, SHAP results reveal a single dominant pattern: \emph{downstream diffusion and diversity are primarily driven by heterogeneity in the upstream intellectual environment}. Features related to volume, variability, and breadth (cited-paper weight, number of cited papers, geometric dispersion) consistently promote exogenous conceptual radiation, while indicators of endogenous specialisation exhibit weak or negligible effects. This confirms that heterogeneous, well-connected citation foundations are the strongest predictors of future conceptual reach.

\subsection{Cross-domain diffusion robustness}
\label{sec:cross_domain}

Table~\ref{tab:metrics_summary} establishes the quantum-computing benchmark under the original stratified split and Optuna-tuned models. To test whether the endo/exo asymmetry persists elsewhere, we replicated the forecasting protocol on robotics, advanced materials, and neuro implants using the OpenAlex validation subsample and a 2022--2023 focal-year holdout (Section~\ref{sec:methods}).

\begin{table}[htbp]
\centering
\caption{Test $R^2$ on the OpenAlex validation subsample (focal years 2022--2023, fixed hyperparameters, \emph{unnormalised} citation counts). The quantum-computing row is \emph{not} comparable to Table~\ref{tab:metrics_summary}: that benchmark uses growth-normalised targets, a stratified 80/20 split, and Optuna-tuned models (endo.\ count $R^2_{\text{test}} \approx 0.02$).}
\small
\begin{tabular}{lrrrr}
\toprule
Domain & Endo.\ count & Exo.\ count & Ratio endo & Citing entropy \\
\midrule
Quantum computer   & 0.47 & 0.60 & 0.08 & 0.43 \\
Robotics           & 0.57 & 0.74 & 0.08 & 0.42 \\
Advanced materials & 0.59 & 0.76 & 0.21 & 0.57 \\
Neuro implants     & 0.83 & 0.87 & 0.21 & 0.45 \\
\bottomrule
\end{tabular}
\label{tab:comparative_r2}
\end{table}

\begin{figure}[htbp]
\centering
\includegraphics[width=0.9\linewidth]{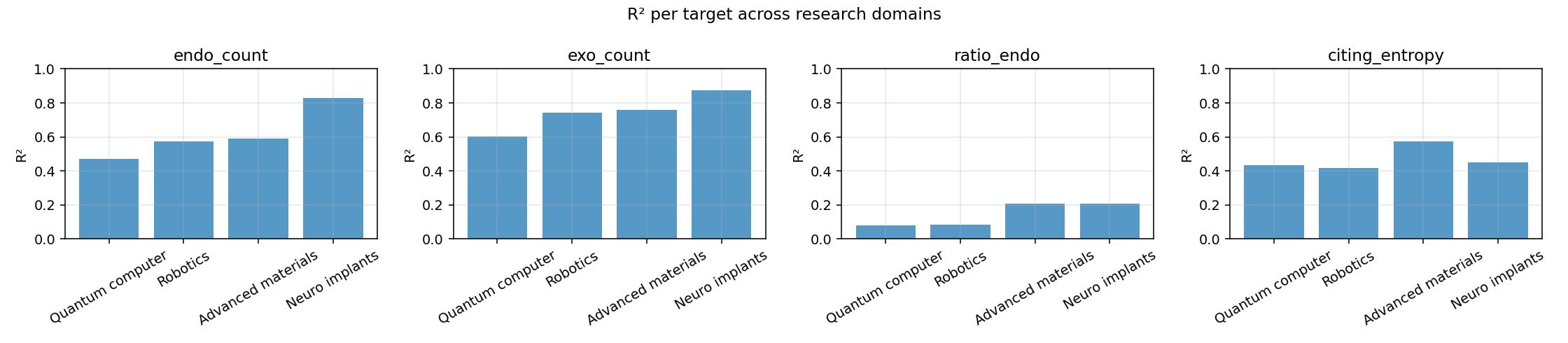}
\caption{Test $R^2$ by regression target across four validation domains.}
\label{fig:comparative_r2}
\end{figure}

Three patterns recur across domains (all on unnormalised counts). First, \emph{exogenous count} remains the most reliably predictable target in every field ($R^2_{\text{test}} \approx 0.60$--$0.87$), reinforcing the claim that cross-pair recombination in citing papers is structurally organised. Second, \emph{endo\_count} is modestly below exo in quantum computing and robotics on this protocol ($R^2_{\text{test}} \approx 0.47$--$0.59$ vs.\ $0.60$--$0.74$) but rises sharply for neuro implants ($0.83$), consistent with a mature biomedical corpus where focal-pair self-reinforcement is denser and more regular; the near-zero endogenous $R^2$ in Table~\ref{tab:metrics_summary} is specific to growth-normalised targets and does not carry over to the comparative holdout. Third, \emph{ratio\_endo} stays low everywhere (typical values $\sim 10^{-3}$; $R^2_{\text{test}} \le 0.21$), so this target should be interpreted cautiously. \emph{Citing entropy} is stable at $R^2_{\text{test}} \approx 0.42$--$0.45$ for quantum computing, robotics, and neuro implants, with a higher value for advanced materials ($0.57$), plausibly reflecting the broader multi-seed upstream heterogeneity of that corpus. Full MAE and RMSE are reported in Appendix~\ref{app:comparative_metrics}. These within-domain replications do not replace the normalized, Optuna-tuned quantum benchmark, but they show that exo-dominated predictability is a cross-field regularity rather than a quantum-computing artefact.

\subsection{Case Studies}
\label{sec:casestudies}
\begin{figure}[h!]
    \centering
    \includegraphics[width=\linewidth]{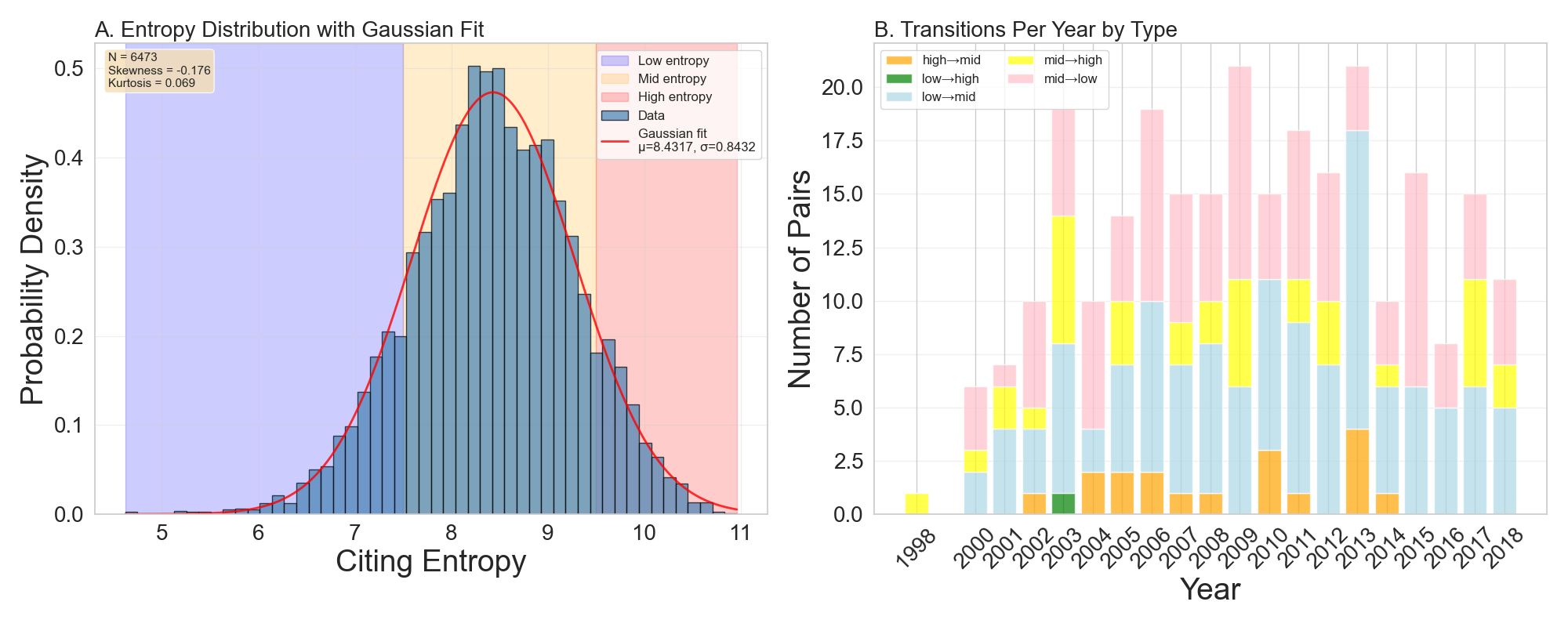}
    \caption{\textbf{Entropy Distribution and Temporal Transition Patterns.}
    {\bf A.} Empirical entropy distribution with Gaussian fit, quartile-based category thresholds, and stable shape across years.  
    {\bf B.} Annual entropy-category transitions, counting only shifts occurring within $\leq 3$ years and exceeding one standard deviation. Coloured stacks show transition types.
    }
    \label{fig:entropy_transition_figure}
\end{figure}

Entropy provides a compact descriptor of how broadly a concept pair diffuses across scientific communities. Its Gaussian distribution (Figure~\ref{fig:entropy_transition_figure}A) yields low-, mid-, and high-entropy regimes corresponding to specialised, field-core, and cross-domain diffusion (for the sake of simplicity, we set low-medium and medium-high thresholds manually to respectively $7.5$ and $9.5$, roughly corresponding to one standard deviation around the mean). Three stable diffusion regimes appear clearly: low-entropy pairs, such as \emph{charge × charge qubit} or \emph{superconductivity × phase qubit}, remain confined to specialised engineering niches; mid-entropy pairs, including \emph{quantum error correction × quantum computer} and \emph{quantum network × QKD}, diffuse across the quantum-information core without crossing field boundaries; and high-entropy pairs, such as \emph{computer science × quantum computer} or \emph{qubit × quantum computer}, spread across algorithms, hardware, condensed matter, and photonics, acting as anchors for cross-domain expansion. 

Our transition-detection procedure isolates only those entropy shifts that are both substantial and temporally meaningful, using two tunable criteria, i.e., {\it category change}, a {\it maximum 3 year inter-observation gap}, and a {\it minimum change of one standard deviation} (to avoid capturing multiple small switches across a threshold). Under these settings, only a limited number of concept pairs undergo significant transitions every year  (Figure~\ref{fig:entropy_transition_figure}B), and these shifts are directly interpretable. Upward transitions typically reflect rapid conceptual expansion (mid → high), while downward transitions signal contraction dynamics such as convergence, technological standardisation, or the abandonment of speculative pathways (high → mid or mid → low). Together, these patterns show that most pairs remain stable, while a small subset exhibits decisive and explainable changes each year in diffusion breadth. 

Sharp entropy increases over 1–3 years mark episodes of rapid conceptual expansion. For instance, \emph{Cluster state × multipartite entanglement} rises from $\approx 7$ to $\geq 9$ over the period 2001-2003, with the recognition of measurement-based quantum computation. \emph{Quantum walk × statistical physics} jumps from $\approx 6.8$ to $\geq 9$ (2002–2004) as quantum walks become a unifying framework for algorithms, transport models, and condensed-matter analogues. \emph{Superconducting QC × phase qubit} increases from $\approx 7.1$ to $\geq 9$ (2000–2003) following the first coherent-oscillation demonstrations, which catalyse expansion into circuit-QED, decoherence studies, and cryogenic engineering. These events represent widening of the scientific search space and the formation of new conceptual frontiers. Entropy collapses provide the complementary pattern: contraction of conceptual reach. \emph{Quantum walk × quantum algorithm} decreases from $\approx 8.4$ to $\approx 6.9$ (2003–2005) as research consolidates around algorithmic complexity. \emph{Quantum teleportation × QKD} drops from $\approx 9.0$ to $\approx 7.8$ (2001–2003) as QKD becomes the dominant applied framework. \emph{Quantum Turing machine × Turing machine} falls from $\approx 7.9$ to $\approx 5.9$ (2001–2003) as the quantum-circuit model displaces the Turing-machine paradigm. These decreases capture technological standardisation, narrowing theoretical focus, or paradigm displacement.

\vspace{0.5cm}
\section{Discussion \& Implications}
\label{sec:discussion}

\subsection{Key insights}

Our study shows that conceptual evolution in quantum science follows a clear asymmetry between persistence and diffusion, consistent with science-of-science work distinguishing cumulative reinforcement from cross-domain recombination (\cite{uzzi_atypical_2013}). In the primary quantum-computing benchmark, once overall growth is accounted for, endogenous reinforcement carries little predictive information on the future influence or diffusion of a concept pair. By contrast, exogenous diffusion and diversity-related outcomes, particularly entropy, follow stable and interpretable regularities: concept pairs embedded in heterogeneous upstream citation environments---with high dispersion and high entropy---are substantially more likely to diffuse across domains (\cite{wu_large_2019,shi_surprising_2023}).

Comparative validation (Section~\ref{sec:cross_domain}) qualifies this picture. In the primary quantum-computing benchmark, endogenous counts are effectively unpredictable after normalisation ($R^2_{\text{test}} \approx 0.02$). Across replications, exogenous diffusion remains the dominant signal in every field, yet endogenous counts become strongly predictable for neuro implants ($R^2_{\text{test}} = 0.83$) while remaining far below exogenous performance in quantum computing and robotics. The quantum-computing asymmetry---endogenous reinforcement negligible relative to exogenous diffusion---therefore does not extend uniformly: mature biomedical corpora can exhibit structured self-reinforcement that the quantum-computing testbed obscures. Examining additional domains should reveal further heterogeneity in how persistence and diffusion co-evolve.

In particular, entropy provides a compact and theoretically meaningful lens on these dynamics. As illustrated by the case studies (Section \ref{sec:casestudies}), sharp entropy expansions reliably mark moments of conceptual opening and the emergence of new research frontiers, while entropy collapses signal consolidation, technological lock-in, or paradigm displacement (\cite{veugelers_scientific_2019}). These patterns reinforce the view of science as a modular yet permeable system in which long-term influence depends less on early self-reinforcement than on sustained cross-community recombination (\cite{fortunato_science_2018}).

Crucially, these structural signals are detectable early: concept pairs that display distinctive upstream diversity profiles later undergo broad downstream diffusion. This establishes diffusion-oriented metrics---particularly entropy and upstream heterogeneity---as actionable signals for regime shift detection and early forecasting.

\subsection{Limitations}

Several limitations qualify these findings. The analysis relies on OpenAlex concept annotations, which may introduce noise at fine semantic resolutions, and on yearly temporal aggregation, which can smooth short-lived dynamics. The focus on concept pairs is a deliberate first-order projection of the underlying co-occurrence hypergraph and excludes higher-order conceptual structures (concept triples, higher-arity simplices) that may capture additional aspects of scientific change.

A further temporal limitation concerns right-censoring of the downstream window. The five-year diffusion horizon ($t{+}1, \ldots, t{+}5$) is fully observable for focal years up to $2018$, but only partially observed for focal years $2019$--$2023$, since the OpenAlex snapshot used here does not yet contain the full citation tail of the most recent cohorts. The primary quantum-computing benchmark (Table~\ref{tab:metrics_summary}) therefore uses only censoring-free focal years $1996$--$2018$; pairs in the censored region are excluded from that fit rather than included with underestimated targets. Comparative validation (Section~\ref{sec:cross_domain}) deliberately holds out recent focal years 2022--2023 and should be read as an early-window protocol on unnormalised counts (Appendix~\ref{app:censoring_compare}). Corpus-level statistics for 1990--2023 and cohort definitions are summarised in Appendix~\ref{app:censoring}.

Methodologically, normalising out global growth strengthens interpretability but limits inference about absolute impact or citation volumes. The primary quantum-computing results (Table~\ref{tab:metrics_summary}) use stratified splits and Optuna-tuned models on normalised targets; the comparative validation in Section~\ref{sec:cross_domain} replicates the same feature set on robotics, advanced materials, and neuro implants using an OpenAlex validation subsample and fixed hyperparameters, and confirms that exogenous diffusion remains the strongest predictable signal across fields while endogenous predictability varies by domain. Absolute $R^2$ values in the comparative table are not directly comparable to Table~\ref{tab:metrics_summary} because of these protocol differences. Artificial intelligence and machine learning remain an important extension for future work, as do cross-domain train-on-one-field / test-on-another experiments.

\subsection{Outlook}
These results open several avenues for future research and application. Extending the framework across domains -- in particular comparative replications on artificial intelligence / machine learning and on biotechnology subtrees of OpenAlex -- would enable comparison of diffusion regimes in fields with different structures and growth dynamics. Moving beyond pairwise co-occurrence toward genuinely higher-order representations (hypergraphs, simplicial complexes, motif-based descriptors) is a natural next step that should preserve the upstream--downstream distinction while accommodating $k$-ary conceptual combinations. Finer temporal resolution, alternative semantic representations, and integration of \emph{non-bibliometric} signals -- patent filings (e.g., USPTO/EPO), funding flows (e.g., CORDIS, NSF), and preprint activity (e.g., arXiv) -- would broaden the evidentiary base beyond peer-reviewed publications and may further improve early detection of conceptual transitions. Integrating entropy-based indicators with neural or graph-embedding approaches offers a promising direction for enhancing sensitivity while preserving interpretability.

From an applied perspective, these findings call for a substantive rethinking of how scientific and technological change is anticipated. Forecasting tools that prioritise endogenous persistence risk systematically overlooking emerging but weakly consolidated ideas, a limitation long noted in innovation and policy studies emphasising path dependence and lock-in (\cite{arthur_nature_2009,schot_three_2018}). In contrast, diffusion-oriented indicators, such as upstream diversity, entropy, and cross-domain positioning, directly target the mechanisms through which new research fronts form and spread, aligning with evidence that recombination and cross-sectoral uptake drive transformative innovation (\cite{fleming_recombinant_2001,georghiou_handbook_2008}). Embedding such metrics into foresight and evaluation pipelines would enable earlier detection of transformative trajectories, clearer differentiation between consolidation and expansion phases, and more robust guidance for strategic investment and science and technology policy. More broadly, quantifying the tension between recombination-driven expansion and convergence-driven consolidation provides a policy-relevant framework for understanding how scientific fields evolve, stabilise, and periodically reorganise.
\vspace{0.5cm}
\section{Conclusion}
\label{sec:conclusion}

Using quantum computing as a testbed, this study shows that conceptual evolution is shaped by a clear asymmetry in that field: endogenous reinforcement mainly reflects proportional growth, while exogenous diffusion and entropy follow predictable patterns rooted in upstream heterogeneity. Comparative replications indicate that this pattern is not universal---endogenous counts are strongly predictable in neuro implants but not in quantum computing---so broader domain coverage is needed to map how diffusion regimes differ across scientific areas. By linking upstream citation lineages to downstream propagation, the framework captures how concept pairs spread beyond local communities. Entropy and citation breadth emerge as key drivers of cross-domain uptake, with expansions marking new conceptual frontiers and collapses signalling convergence. Because the upstream signatures of these shifts are detectable early, the approach provides a concise and scalable basis for anticipatory scientometrics and the identification of emerging research fronts.

\paragraph{Data \& Source code:} All code and the manuscript sources are openly available at \url{https://github.com/wazaahhh/breakthroughs-diffusion}. The analysis is built entirely on the open OpenAlex corpus (\url{https://openalex.org}).

\paragraph{Acknowledgments:}
T.C. and T.M. acknowledge funding from armasuisse Science + Technology. D.D. acknowledges support from Open Quantum Institute.

\vspace{0.5cm}

\printbibliography

@inproceedings{dolamic_automated_2024,
	address = {Changchun, China},
	title = {Automated {Identification} of {Emerging} {Technologies}: {Open} {Data} {Approach}},
	author = {Dolamic, Ljiljana and Jang-Jaccard, Julian and Mermoud, Alain and Lenders, Vincent},
	year = {2024},
}

@article{strumsky_complexity_2010,
	title = {Complexity and the productivity of innovation},
	volume = {27},
	copyright = {Copyright © 2010 John Wiley \& Sons, Ltd.},
	issn = {1099-1743},
	url = {https://onlinelibrary.wiley.com/doi/abs/10.1002/sres.1057},
	doi = {10.1002/sres.1057},
	language = {en},
	number = {5},
	urldate = {2025-11-14},
	journal = {Systems Research and Behavioral Science},
	author = {Strumsky, Deborah and Lobo, José and Tainter, Joseph A.},
	year = {2010},
	pages = {496--509},
}

@article{pan_memory_2018,
	title = {The memory of science: {Inflation}, myopia, and the knowledge network},
	volume = {12},
	issn = {1751-1577},
	shorttitle = {The memory of science},
	url = {https://www.sciencedirect.com/science/article/pii/S1751157717303139},
	doi = {10.1016/j.joi.2018.06.005},
	number = {3},
	urldate = {2025-11-14},
	journal = {Journal of Informetrics},
	author = {Pan, Raj K. and Petersen, Alexander M. and Pammolli, Fabio and Fortunato, Santo},
	month = aug,
	year = {2018},
	pages = {656--678},
}

@article{chen_science_2017,
	title = {Science mapping: a systematic review of the literature},
	volume = {2},
	number = {2},
	journal = {Journal of data and information science},
	author = {Chen, Chaomei},
	year = {2017},
}

@book{kuhn_structure_1997,
	title = {The structure of scientific revolutions},
	volume = {962},
	publisher = {University of Chicago press Chicago},
	author = {Kuhn, Thomas S},
	year = {1997},
}

@article{veugelers_scientific_2019,
	title = {Scientific novelty and technological impact},
	volume = {48},
	issn = {0048-7333},
	url = {https://www.sciencedirect.com/science/article/pii/S0048733319300459},
	doi = {10.1016/j.respol.2019.01.019},
	number = {6},
	urldate = {2025-11-14},
	journal = {Research Policy},
	author = {Veugelers, Reinhilde and Wang, Jian},
	month = jul,
	year = {2019},
	pages = {1362--1372},
}

@article{youn_invention_2015,
	title = {Invention as a combinatorial process: evidence from {US} patents},
	volume = {12},
	shorttitle = {Invention as a combinatorial process},
	url = {https://royalsocietypublishing.org/doi/full/10.1098/rsif.2015.0272},
	doi = {10.1098/rsif.2015.0272},
	number = {106},
	urldate = {2025-11-14},
	journal = {Journal of The Royal Society Interface},
	author = {Youn, Hyejin and Strumsky, Deborah and Bettencourt, Luis M. A. and Lobo, José},
	month = may,
	year = {2015},
	note = {Publisher: Royal Society},
	pages = {20150272},
}

@article{shi_surprising_2023,
	title = {Surprising combinations of research contents and contexts are related to impact and emerge with scientific outsiders from distant disciplines},
	volume = {14},
	copyright = {2023 The Author(s)},
	issn = {2041-1723},
	url = {https://www.nature.com/articles/s41467-023-36741-4},
	doi = {10.1038/s41467-023-36741-4},
	language = {en},
	number = {1},
	urldate = {2025-11-12},
	journal = {Nature Communications},
	author = {Shi, Feng and Evans, James},
	month = mar,
	year = {2023},
	note = {Publisher: Nature Publishing Group},
	pages = {1641},
}

@article{min_predicting_2021,
	title = {Predicting scientific breakthroughs based on knowledge structure variations},
	volume = {164},
	issn = {0040-1625},
	url = {https://www.sciencedirect.com/science/article/pii/S0040162520313287},
	doi = {10.1016/j.techfore.2020.120502},
	language = {en},
	urldate = {2022-11-11},
	journal = {Technological Forecasting and Social Change},
	author = {Min, Chao and Bu, Yi and Sun, Jianjun},
	month = mar,
	year = {2021},
	pages = {120502},
}

@article{wu_large_2019,
	title = {Large teams develop and small teams disrupt science and technology},
	volume = {566},
	copyright = {2019 The Author(s), under exclusive licence to Springer Nature Limited},
	issn = {1476-4687},
	url = {https://www.nature.com/articles/s41586-019-0941-9},
	doi = {10.1038/s41586-019-0941-9},
	language = {en},
	number = {7744},
	urldate = {2022-10-11},
	journal = {Nature},
	author = {Wu, Lingfei and Wang, Dashun and Evans, James A.},
	month = feb,
	year = {2019},
	note = {Number: 7744
Publisher: Nature Publishing Group},
	pages = {378--382},
}

@article{wagner_international_2019,
	title = {International research collaboration: {Novelty}, conventionality, and atypicality in knowledge recombination},
	volume = {48},
	issn = {0048-7333},
	shorttitle = {International research collaboration},
	url = {https://www.sciencedirect.com/science/article/pii/S0048733319300046},
	doi = {10.1016/j.respol.2019.01.002},
	number = {5},
	urldate = {2025-11-14},
	journal = {Research Policy},
	author = {Wagner, Caroline S. and Whetsell, Travis A. and Mukherjee, Satyam},
	month = jun,
	year = {2019},
	pages = {1260--1270},
}

@book{arthur_nature_2009,
	title = {The {Nature} of {Technology}: {What} {It} {Is} and {How} {It} {Evolves}},
	isbn = {978-1-4391-6578-2},
	shorttitle = {The {Nature} of {Technology}},
	language = {en},
	publisher = {Simon and Schuster},
	author = {Arthur, W. Brian},
	month = aug,
	year = {2009},
}

@article{percia_david_measuring_2023,
	title = {Measuring security development in information technologies: {A} scientometric framework using {arXiv} e-prints},
	volume = {188},
	issn = {00401625},
	shorttitle = {Measuring security development in information technologies},
	url = {https://linkinghub.elsevier.com/retrieve/pii/S004016252300001X},
	doi = {10.1016/j.techfore.2023.122316},
	language = {en},
	urldate = {2025-10-26},
	journal = {Technological Forecasting and Social Change},
	author = {Percia David, Dimitri and Maréchal, Loïc and Lacube, William and Gillard, Sébastien and Tsesmelis, Michael and Maillart, Thomas and Mermoud, Alain},
	month = mar,
	year = {2023},
	pages = {122316},
}

@article{wang_measuring_2008,
	title = {Measuring the preferential attachment mechanism in citation networks},
	volume = {387},
	issn = {0378-4371},
	url = {https://www.sciencedirect.com/science/article/pii/S0378437108003208},
	doi = {10.1016/j.physa.2008.03.017},
	number = {18},
	urldate = {2025-11-14},
	journal = {Physica A: Statistical Mechanics and its Applications},
	author = {Wang, Mingyang and Yu, Guang and Yu, Daren},
	month = jul,
	year = {2008},
	pages = {4692--4698},
}

@article{lundberg_local_2020,
	title = {From local explanations to global understanding with explainable {AI} for trees},
	volume = {2},
	copyright = {2020 The Author(s), under exclusive licence to Springer Nature Limited},
	issn = {2522-5839},
	url = {https://www.nature.com/articles/s42256-019-0138-9},
	doi = {10.1038/s42256-019-0138-9},
	language = {en},
	number = {1},
	urldate = {2025-11-14},
	journal = {Nature Machine Intelligence},
	author = {Lundberg, Scott M. and Erion, Gabriel and Chen, Hugh and DeGrave, Alex and Prutkin, Jordan M. and Nair, Bala and Katz, Ronit and Himmelfarb, Jonathan and Bansal, Nisha and Lee, Su-In},
	month = jan,
	year = {2020},
	note = {Publisher: Nature Publishing Group},
	pages = {56--67},
}

@inproceedings{lundberg_unified_2017,
	title = {A {Unified} {Approach} to {Interpreting} {Model} {Predictions}},
	volume = {30},
	url = {https://proceedings.neurips.cc/paper/2017/hash/8a20a8621978632d76c43dfd28b67767-Abstract.html},
	urldate = {2025-11-14},
	booktitle = {Advances in {Neural} {Information} {Processing} {Systems}},
	publisher = {Curran Associates, Inc.},
	author = {Lundberg, Scott M and Lee, Su-In},
	year = {2017},
}

@inproceedings{ke_lightgbm_2017,
	title = {{LightGBM}: {A} {Highly} {Efficient} {Gradient} {Boosting} {Decision} {Tree}},
	volume = {30},
	shorttitle = {{LightGBM}},
	url = {https://proceedings.neurips.cc/paper/2017/hash/6449f44a102fde848669bdd9eb6b76fa-Abstract.html},
	urldate = {2025-11-14},
	booktitle = {Advances in {Neural} {Information} {Processing} {Systems}},
	publisher = {Curran Associates, Inc.},
	author = {Ke, Guolin and Meng, Qi and Finley, Thomas and Wang, Taifeng and Chen, Wei and Ma, Weidong and Ye, Qiwei and Liu, Tie-Yan},
	year = {2017},
}

@inproceedings{akiba_optuna_2019,
	address = {New York, NY, USA},
	series = {{KDD} '19},
	title = {Optuna: {A} {Next}-generation {Hyperparameter} {Optimization} {Framework}},
	isbn = {978-1-4503-6201-6},
	shorttitle = {Optuna},
	url = {https://dl.acm.org/doi/10.1145/3292500.3330701},
	doi = {10.1145/3292500.3330701},
	urldate = {2025-11-14},
	booktitle = {Proceedings of the 25th {ACM} {SIGKDD} {International} {Conference} on {Knowledge} {Discovery} \& {Data} {Mining}},
	publisher = {Association for Computing Machinery},
	author = {Akiba, Takuya and Sano, Shotaro and Yanase, Toshihiko and Ohta, Takeru and Koyama, Masanori},
	month = jul,
	year = {2019},
	pages = {2623--2631},
}

@article{newman_structure_2001,
	title = {The structure of scientific collaboration networks},
	volume = {98},
	url = {https://www.pnas.org/doi/full/10.1073/pnas.98.2.404},
	doi = {10.1073/pnas.98.2.404},
	number = {2},
	urldate = {2025-11-14},
	journal = {Proceedings of the National Academy of Sciences},
	author = {Newman, M. E. J.},
	month = jan,
	year = {2001},
	note = {Publisher: Proceedings of the National Academy of Sciences},
	pages = {404--409},
}

@inproceedings{enduri_does_2015,
	title = {Does {Diversity} of {Papers} {Affect} {Their} {Citations}? {Evidence} from {American} {Physical} {Society} {Journals}},
	shorttitle = {Does {Diversity} of {Papers} {Affect} {Their} {Citations}?},
	url = {https://ieeexplore.ieee.org/abstract/document/7400609},
	doi = {10.1109/SITIS.2015.60},
	urldate = {2025-11-14},
	booktitle = {2015 11th {International} {Conference} on {Signal}-{Image} {Technology} \& {Internet}-{Based} {Systems} ({SITIS})},
	author = {Enduri, Murali Krishna and Reddy, I. Vinod and Jolad, Shivakumar},
	month = nov,
	year = {2015},
	pages = {505--511},
}

@article{rzhetsky_choosing_2015,
	title = {Choosing experiments to accelerate collective discovery},
	volume = {112},
	url = {https://www.pnas.org/doi/abs/10.1073/pnas.1509757112},
	doi = {10.1073/pnas.1509757112},
	number = {47},
	urldate = {2025-11-14},
	journal = {Proceedings of the National Academy of Sciences},
	author = {Rzhetsky, Andrey and Foster, Jacob G. and Foster, Ian T. and Evans, James A.},
	month = nov,
	year = {2015},
	note = {Publisher: Proceedings of the National Academy of Sciences},
	pages = {14569--14574},
}

@article{salatino_how_2017,
	title = {How are topics born? {Understanding} the research dynamics preceding the emergence of new areas},
	volume = {3},
	issn = {2376-5992},
	shorttitle = {How are topics born?},
	url = {https://peerj.com/articles/cs-119},
	doi = {10.7717/peerj-cs.119},
	language = {en},
	urldate = {2025-11-14},
	journal = {PeerJ Computer Science},
	author = {Salatino, Angelo A. and Osborne, Francesco and Motta, Enrico},
	month = jun,
	year = {2017},
	note = {Publisher: PeerJ Inc.},
	pages = {e119},
}

@book{georghiou_handbook_2008,
	title = {The handbook of technology foresight: concepts and practice},
	publisher = {Edward Elgar Publishing},
	author = {Georghiou, Luke},
	year = {2008},
}

@article{schot_three_2018,
	title = {Three frames for innovation policy: {R}\&{D}, systems of innovation and transformative change},
	volume = {47},
	issn = {0048-7333},
	shorttitle = {Three frames for innovation policy},
	url = {https://www.sciencedirect.com/science/article/pii/S0048733318301987},
	doi = {10.1016/j.respol.2018.08.011},
	number = {9},
	urldate = {2025-11-12},
	journal = {Research Policy},
	author = {Schot, Johan and Steinmueller, W. Edward},
	month = nov,
	year = {2018},
	pages = {1554--1567},
}

@book{saichev_theory_2009,
	edition = {1st Edition.},
	title = {Theory of {Zipf}'s {Law} and {Beyond} ({Lecture} {Notes} in {Economics} and {Mathematical} {Systems})},
	isbn = {3-642-02945-0},
	url = {http://www.worldcat.org/isbn/3642029450},
	publisher = {Springer},
	author = {Saichev, Alexander I. and Malevergne, Yannick and Sornette, Didier},
	month = nov,
	year = {2009},
	note = {Published: Paperback},
}

@article{maillart_empirical_2008,
	title = {Empirical {Tests} of {Zipf}'s {Law} {Mechanism} in {Open} {Source} {Linux} {Distribution}},
	volume = {101},
	url = {https://link.aps.org/doi/10.1103/PhysRevLett.101.218701},
	doi = {10.1103/PhysRevLett.101.218701},
	number = {21},
	urldate = {2021-03-28},
	journal = {Physical Review Letters},
	author = {Maillart, T. and Sornette, D. and Spaeth, S. and von Krogh, G.},
	month = nov,
	year = {2008},
	note = {Publisher: American Physical Society},
	pages = {218701},
}

@article{fleming_recombinant_2001,
	title = {Recombinant {Uncertainty} in {Technological} {Search}},
	volume = {47},
	issn = {0025-1909},
	url = {https://pubsonline.informs.org/doi/abs/10.1287/mnsc.47.1.117.10671},
	doi = {10.1287/mnsc.47.1.117.10671},
	number = {1},
	urldate = {2025-11-07},
	journal = {Management Science},
	author = {Fleming, Lee},
	month = jan,
	year = {2001},
	note = {Publisher: INFORMS},
	pages = {117--132},
}

@article{krenn_forecasting_2023,
	title = {Forecasting the future of artificial intelligence with machine learning-based link prediction in an exponentially growing knowledge network},
	volume = {5},
	copyright = {2023 The Author(s)},
	issn = {2522-5839},
	url = {https://www.nature.com/articles/s42256-023-00735-0},
	doi = {10.1038/s42256-023-00735-0},
	language = {en},
	number = {11},
	urldate = {2025-10-24},
	journal = {Nature Machine Intelligence},
	author = {Krenn, Mario and Buffoni, Lorenzo and Coutinho, Bruno and Eppel, Sagi and Foster, Jacob Gates and Gritsevskiy, Andrew and Lee, Harlin and Lu, Yichao and Moutinho, João P. and Sanjabi, Nima and Sonthalia, Rishi and Tran, Ngoc Mai and Valente, Francisco and Xie, Yangxinyu and Yu, Rose and Kopp, Michael},
	month = nov,
	year = {2023},
	note = {Publisher: Nature Publishing Group},
	pages = {1326--1335},
}

@article{gu_forecasting_2025,
	title = {Forecasting high-impact research topics via machine learning on evolving knowledge graphs},
	volume = {6},
	issn = {2632-2153},
	url = {https://doi.org/10.1088/2632-2153/add6ef},
	doi = {10.1088/2632-2153/add6ef},
	language = {en},
	number = {2},
	urldate = {2025-10-16},
	journal = {Machine Learning: Science and Technology},
	author = {Gu, Xuemei and Krenn, Mario},
	month = may,
	year = {2025},
	note = {Publisher: IOP Publishing},
	pages = {025041},
}

@article{jost_entropy_2006,
	title = {Entropy and diversity},
	volume = {113},
	issn = {1600-0706},
	url = {https://onlinelibrary.wiley.com/doi/abs/10.1111/j.2006.0030-1299.14714.x},
	doi = {10.1111/j.2006.0030-1299.14714.x},
	language = {en},
	number = {2},
	urldate = {2022-11-12},
	journal = {Oikos},
	author = {Jost, Lou},
	year = {2006},
	note = {\_eprint: https://onlinelibrary.wiley.com/doi/pdf/10.1111/j.2006.0030-1299.14714.x},
	pages = {363--375},
}

@article{sinatra_quantifying_2016,
	title = {Quantifying the evolution of individual scientific impact},
	volume = {354},
	url = {https://www.science.org/doi/10.1126/science.aaf5239},
	doi = {10.1126/science.aaf5239},
	number = {6312},
	urldate = {2022-10-11},
	journal = {Science},
	author = {Sinatra, Roberta and Wang, Dashun and Deville, Pierre and Song, Chaoming and Barabási, Albert-László},
	month = nov,
	year = {2016},
	note = {Publisher: American Association for the Advancement of Science},
	pages = {aaf5239},
}

@article{uzzi_atypical_2013,
	title = {Atypical {Combinations} and {Scientific} {Impact}},
	volume = {342},
	url = {https://www.science.org/doi/full/10.1126/science.1240474},
	doi = {10.1126/science.1240474},
	number = {6157},
	urldate = {2025-11-07},
	journal = {Science},
	author = {Uzzi, Brian and Mukherjee, Satyam and Stringer, Michael and Jones, Ben},
	month = oct,
	year = {2013},
	note = {Publisher: American Association for the Advancement of Science},
	pages = {468--472},
}

@article{krenn_predicting_2020,
	title = {Predicting research trends with semantic and neural networks with an application in quantum physics},
	volume = {117},
	url = {https://www.pnas.org/doi/full/10.1073/pnas.1914370116},
	doi = {10.1073/pnas.1914370116},
	number = {4},
	urldate = {2025-10-24},
	journal = {Proceedings of the National Academy of Sciences},
	author = {Krenn, Mario and Zeilinger, Anton},
	month = jan,
	year = {2020},
	note = {Publisher: Proceedings of the National Academy of Sciences},
	pages = {1910--1916},
}

@article{fortunato_science_2018,
	title = {Science of science},
	volume = {359},
	url = {https://www.science.org/doi/full/10.1126/science.aao0185},
	doi = {10.1126/science.aao0185},
	number = {6379},
	urldate = {2022-11-12},
	journal = {Science},
	author = {Fortunato, Santo and Bergstrom, Carl T. and Börner, Katy and Evans, James A. and Helbing, Dirk and Milojević, Staša and Petersen, Alexander M. and Radicchi, Filippo and Sinatra, Roberta and Uzzi, Brian and Vespignani, Alessandro and Waltman, Ludo and Wang, Dashun and Barabási, Albert-László},
	month = mar,
	year = {2018},
	note = {Publisher: American Association for the Advancement of Science},
	pages = {eaao0185},
}

@misc{priem_openalex_2022,
	title = {{OpenAlex}: {A} fully-open index of scholarly works, authors, venues, institutions, and concepts},
	shorttitle = {{OpenAlex}},
	url = {http://arxiv.org/abs/2205.01833},
	doi = {10.48550/arXiv.2205.01833},
	urldate = {2022-10-12},
	publisher = {arXiv},
	author = {Priem, Jason and Piwowar, Heather and Orr, Richard},
	month = jun,
	year = {2022},
	note = {arXiv:2205.01833 [cs]},
}

@article{zhang_predicting_2024,
	title = {Predicting citation impact of academic papers across research areas using multiple models and early citations},
	volume = {129},
	journal = {Scientometrics},
	author = {Zhang, Fang and Wu, Shengli},
	year = {2024},
	note = {Publisher: Springer},
	pages = {4137--4166},
}

@article{mistele_predicting_2019,
	title = {Predicting authors’ citation counts and h-indices with a neural network},
	volume = {120},
	journal = {Scientometrics},
	author = {Mistele, Tobias and Price, Tom and Hossenfelder, Sabine},
	year = {2019},
	note = {Publisher: Springer},
	pages = {87--104},
}

@article{gui_technology_2021,
	title = {Technology forecasting using deep learning neural network: {Taking} the case of robotics},
	volume = {9},
	journal = {Ieee Access},
	author = {Gui, Meizeng and Xu, Xueguo},
	year = {2021},
	note = {Publisher: IEEE},
	pages = {53306--53316},
}

@article{hu_technology_2022,
	title = {Technology topic identification and trend prediction of new energy vehicle using {LDA} modeling},
	volume = {2022},
	doi = {https://doi.org/10.1155/2022/9373911},
	number = {1},
	journal = {Complexity},
	author = {Hu, Renjie and Ma, Wencong and Lin, Weiqiang and Chen, Xiude and Zhong, Zuchang and Zeng, Chuhong},
	year = {2022},
	note = {Publisher: Wiley Online Library},
	pages = {9373911},
}

@article{preskill_quantum_2018,
	title = {Quantum computing in the {NISQ} era and beyond},
	volume = {2},
	doi = {10.22331/q-2018-08-06-79},
	journal = {Quantum},
	author = {Preskill, John},
	year = {2018},
	pages = {79},
}

@article{li_technology_2022,
	title = {Technology opportunity discovery using deep learning-based text mining and knowledge graphs},
	volume = {177},
	doi = {https://doi.org/10.1016/j.techfore.2022.122161},
	journal = {Technological Forecasting and Social Change},
	author = {Li, H. and Zhang, Y. and Wang, C. and Liu, F.},
	year = {2022},
	pages = {121506},
}



\renewcommand{\thesection}{A.\arabic{section}}
\renewcommand{\thesubsection}{A.\arabic{section}.\arabic{subsection}}

\setcounter{section}{0} 
\setcounter{figure}{0}
\setcounter{table}{0}

\renewcommand{\thefigure}{A\arabic{figure}}
\renewcommand{\thetable}{A\arabic{table}}

\clearpage
\appendix
\section{Right-censoring and focal-year cohorts}
\label{app:censoring}

\subsection{Cohort definitions}
\label{app:cohorts}

The OpenAlex snapshot used here ends in calendar year 2023. For a focal year $t$, downstream targets aggregate citing papers over $(t{+}1,\ldots,t{+}5)$. That window is \emph{fully} observed only when $t \le 2018$; for $t \in \{2019,\ldots,2023\}$, later citing years are missing from the snapshot, so absolute downstream weights and entropies are right-censored (underestimated). Upstream features at $t{-}1$ remain fully observed for all focal years in the corpus.

Table~\ref{tab:cohort_summary} summarises the focal-year cohorts. The primary quantum-computing benchmark (Table~\ref{tab:metrics_summary} in the main text) is fit exclusively on the \textbf{censoring-free} cohort: focal years 1996--2018, 6\,978 pair--years, growth-normalised targets, stratified 80/20 train--test split, and Optuna-tuned LightGBM models. This matches the extraction and training code (\texttt{7\_predict\_task.ipynb}, filter \texttt{base\_year < 2019}).

\begin{table}[htbp]
\caption{Focal-year cohorts for the quantum-computing modelling dataset. Table~\ref{tab:metrics_summary} is fit on the censoring-free cohort only.}
\label{tab:cohort_summary}
\small
\begin{tabular}{lrrrp{3.2cm}}
\toprule
Cohort & Focal years & Pair--years & Window & Role \\
\midrule
Censoring-free (primary) & 1996--2018 & 6,978 & complete $t{+}1,\ldots,t{+}5$ & Table~\ref{tab:metrics_summary} \\
Snapshot-inclusive & 1996--2023 & ---$^\dagger$ & mixed (complete if $t\le 2018$) & robustness (pipeline) \\
Censored only & 2019--2023 & ---$^\dagger$ & partial window only & robustness (pipeline) \\
\bottomrule
\end{tabular}
\end{table}

\noindent$^\dagger$Not present in the archived primary feature matrix; rebuild with \texttt{pipeline.sub117\_run} without \texttt{--max-focal-year~2018}.

\subsection{Relation to the comparative validation protocol}
\label{app:censoring_compare}

Section~\ref{sec:cross_domain} and Appendix~\ref{app:transfer} hold out focal years 2022--2023 on the OpenAlex validation subsample. Those focal years lie in the censored region: the five-year downstream horizon is incomplete in the 2023 snapshot. Comparative $R^2$ values therefore reflect \emph{early} citation structure on unnormalised counts (fixed hyperparameters), not the same complete-window protocol as Table~\ref{tab:metrics_summary}. The two tables are complementary, not directly comparable row-for-row.

\subsection{Robustness to snapshot-inclusive focal years}
\label{app:censoring_robust}

Re-fitting on a snapshot-inclusive cohort ($t \in \{1996,\ldots,2023\}$) or on censored-only years ($t \in \{2019,\ldots,2023\}$) requires rebuilding pair--year features without the focal-year cap (e.g.\ \texttt{python -m pipeline.sub117\_run --subdomain quantum\_computer --min-focal-year 1996} without \texttt{--max-focal-year 2018}). Because the archived primary dataset already excludes $t \ge 2019$, the headline exogenous-count and entropy $R^2_{\text{test}}$ in Table~\ref{tab:metrics_summary} are \emph{not} driven by censored downstream tails. We do not expect qualitative conclusions to change once inclusive-cohort metrics are added; those runs are reserved for the journal sensitivity appendix when the unfiltered feature matrix is regenerated.

\section{Comparative validation across four research domains}
\label{app:transfer}

\subsection{Domain definitions and OpenAlex concept seeds}
\label{app:domains}

Table~\ref{tab:app_domains} lists the four OpenAlex concept subtrees used in the comparative validation study (Section~\ref{sec:cross_domain}). Corpus sizes refer to the OpenAlex validation subsample (Section~\ref{sec:methods}).

\begin{table}[htbp]
\caption{Research domains, concept seeds, and corpus characteristics on the OpenAlex validation subsample. Ann.\ growth: annual corpus growth rate, 1990--2023.}
\small
\begin{tabular}{p{2.6cm}p{4.8cm}rrr}
\toprule
Domain & OpenAlex seed concept(s) & Subtree & Works & Ann.\ growth \\
\midrule
Quantum computer (baseline) & C58053490 (\emph{Quantum computer}, L3) & 3 & 31\,935 & +22.7\,\% \\
Robotics & C34413123 (\emph{Robotics}, L3) & 6 & 51\,443 & +15.2\,\% \\
Advanced materials & C138631740 (Nanomaterials), C110367647 (Metamaterial), C2778414984 (Biomaterial), C88484716 (Smart material) & 9 & 104\,707 & +11.3\,\% \\
Neuro implants & C173201364 (BCI), C2780375056 (Neuromodulation), C197525751 (Neuroprosthetics), C2778542668 (DBS), C2778882171 (Cochlear implant), C2776443511 (Neurostimulation) & 12 & 70\,883 & +9.4\,\% \\
\bottomrule
\end{tabular}
\label{tab:app_domains}
\end{table}

\subsection{Full comparative validation metrics}
\label{app:comparative_metrics}

Table~\ref{tab:app_comparative_full} reports test-set $R^2$, MAE, and RMSE for all four domains and regression targets, evaluated on focal years 2022--2023 with fixed LightGBM hyperparameters on \emph{unnormalised} citation counts. Those focal years lie in the right-censored region of the 2023 snapshot (incomplete five-year downstream windows; Appendix~\ref{app:censoring_compare}). Quantum-computer endogenous $R^2 = 0.47$ here vs.\ $0.018$ in Table~\ref{tab:metrics_summary} further reflects growth normalisation, censoring-free focal years 1996--2018, and a stratified Optuna-tuned split in the primary benchmark, not a data error.

\begin{table}[htbp]
\caption{Comparative validation: test-set metrics by domain and target.}
\small
\setlength{\tabcolsep}{4pt}
\begin{tabular}{llrrrrr}
\toprule
Domain & Target & $R^2$ & MAE & RMSE & $n_{\text{train}}$ & $n_{\text{test}}$ \\
\midrule
Quantum computer & Endo.\ count & 0.470 & 33.17 & 231.34 & 29\,966 & 8\,930 \\
                 & Exo.\ count  & 0.601 & 9\,567.78 & 28\,343.96 & 29\,966 & 8\,930 \\
                 & Ratio endo   & 0.081 & 0.00110 & 0.00182 & 29\,966 & 8\,930 \\
                 & Citing entropy & 0.434 & 0.634 & 0.937 & 29\,966 & 8\,930 \\
\midrule
Robotics & Endo.\ count & 0.572 & 7.14 & 68.32 & 26\,162 & 10\,241 \\
                  & Exo.\ count  & 0.742 & 2\,327.89 & 9\,056.45 & 26\,162 & 10\,241 \\
                  & Ratio endo   & 0.083 & 0.00136 & 0.00232 & 26\,162 & 10\,241 \\
                  & Citing entropy & 0.417 & 0.780 & 1.038 & 26\,162 & 10\,241 \\
\midrule
Advanced materials & Endo.\ count & 0.592 & 11.28 & 132.68 & 107\,065 & 31\,574 \\
                   & Exo.\ count  & 0.760 & 3\,551.51 & 17\,366.65 & 107\,065 & 31\,574 \\
                   & Ratio endo   & 0.208 & 0.00096 & 0.00150 & 107\,065 & 31\,574 \\
                   & Citing entropy & 0.572 & 0.509 & 0.737 & 107\,065 & 31\,574 \\
\midrule
Neuro implants & Endo.\ count & 0.830 & 32.31 & 97.95 & 77\,719 & 16\,072 \\
               & Exo.\ count  & 0.873 & 5\,487.36 & 9\,298.54 & 77\,719 & 16\,072 \\
               & Ratio endo   & 0.209 & 0.00139 & 0.00202 & 77\,719 & 16\,072 \\
               & Citing entropy & 0.449 & 0.535 & 0.789 & 77\,719 & 16\,072 \\
\bottomrule
\end{tabular}
\label{tab:app_comparative_full}
\end{table}

\begin{figure}[htbp]
\centering
\includegraphics[width=0.95\linewidth]{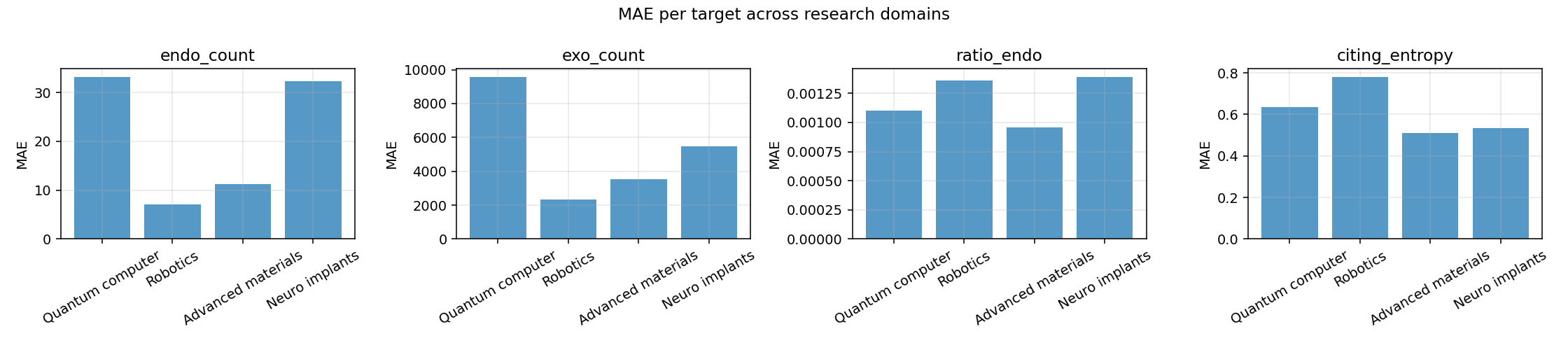}
\caption{Test-set MAE by regression target across the four validation domains.}
\label{fig:app_mae_comparative}
\end{figure}

\begin{figure}[htbp]
\centering
\includegraphics[width=0.95\linewidth]{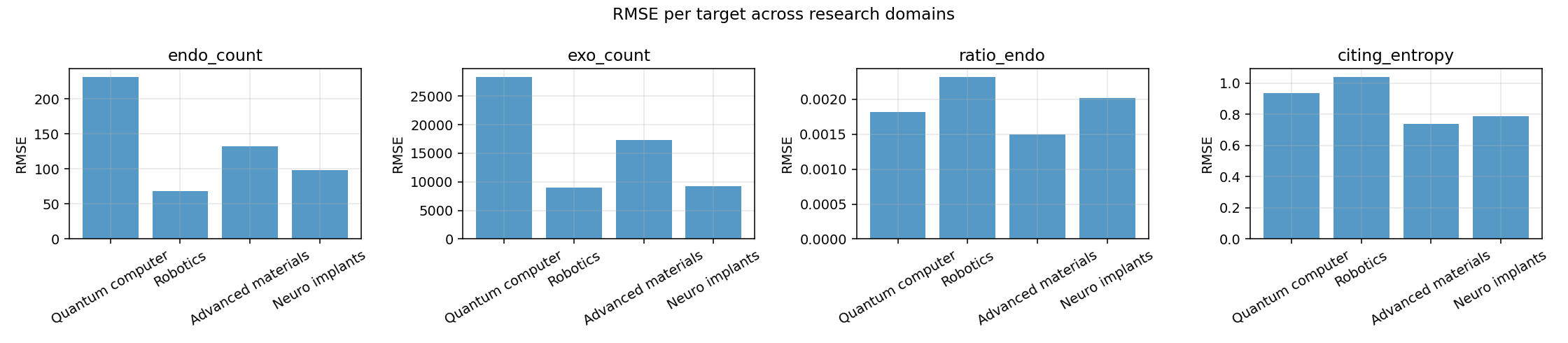}
\caption{Test-set RMSE by regression target across the four validation domains.}
\label{fig:app_rmse_comparative}
\end{figure}

\paragraph{Reproducibility.}
The comparative validation uses the same upstream-feature construction, target definitions, and LightGBM protocol described in Section~\ref{sec:methods}.

\end{document}